\documentclass[times]{elsarticle}
\newcommand\Tr{\hbox{Tr }}
\usepackage{graphicx}
\biboptions{sort&compress}

\title{Auto-correlation Functions and Quantum Fluctuations
of the Transverse Ising Chain by the Quantum Transfer Matrix Method}
\author{Makoto INOUE }

\address{Division of Science, Tokyo Denki University \\
Hatoyama, Sitama, 350-0394, Japan
}

\begin{document}

\begin{abstract}
The Quantum Transfer Matrix method based on the Suzuki-Trotter formulation
is extended to dynamical problems.
The auto-correlation functions of the Transverse Ising chain
are derived by this method.
It is shown that the  Trotter-directional correlation function 
is interpreted as a Matsubara's temperature Green function 
and that the auto-correlation function is given by 
analytic continuation of the Green function.

We propose the Trotter-directional correlation function 
is a new measure of the quantum fluctuation and show
how it works well as a demonstration.

\end{abstract}

\begin{keyword}
Auto-correlation Function,Quantum Fluctuation,
Temperature Green Function,
Suzuki-Trotter Transformation, Quantum Transfer Matrix,
Transverse Ising Chain
\end{keyword}

\maketitle

\section{Introduction}
The Quantum Transfer Matrix(QTM) method is a so powerful tool
to study thermodynamic properties of low-dimensional quantum systems.
It was proposed by Suzuki \cite{MS-QTM1} and was demonstrated by Suzuki and the
present author \cite{MI} for the XY quantum spin chains.
It was widely used to one-dimensional integrable quantum systems 
\cite{Koma, SWA, deVega, AK-ST,  Takahashi, S3U, JKS, Corr-XXZ}
 and to non-integrable systems \cite{MS-MC}.
We extend this method to study a dynamical problem 
by using a Transverse Ising chain as an example in this paper.

\begin{figure}[tbp]
  \begin{center}
    \includegraphics[keepaspectratio=true,height=60mm]{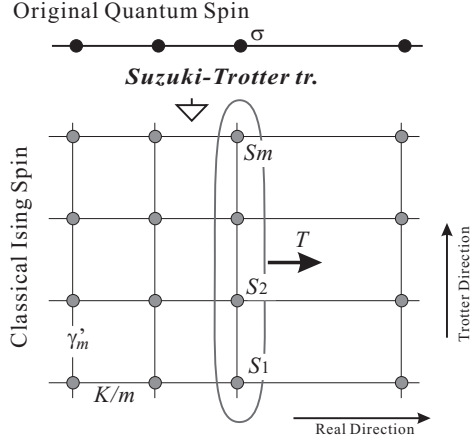}
  \end{center}
  \caption{The original quantum spin ($\sigma$) chain and
  the ST-transformed two-dimensional Ising ($S$) system. 
  The QTM {$\cal T$} transfers 
  along to the real direction. 
  }
  \label{fig-STtr}
\end{figure}

The Suzuki-Trotter (ST) transformation maps a $d$-dimensional quantum
system to a $d+1$-dimensional classical system adding an 
extra Trotter dimension \cite{MS-QTM1,MS-ST,MS-MC,DeRaedt}.
The QTM $\cal T$ transfers along to the real dimension as shown in
fig.\ref{fig-STtr}.
The free energy of the relevant quantum system in the thermodynamic limit
is given by the one maximum eigenvalue $\lambda_{max}$ of $\cal T$
as follows \cite{MI,Koma}.
\begin{equation}\label{QTM}
-\beta f = \lim_{m \to \infty} \log \lambda_{{max}}, 
\qquad \hbox{where} \qquad
{\cal T} |\psi_{max}\rangle_m = \lambda_{max} |\psi_{max}\rangle_m  .
\end{equation}
Where we add a suffix $m$ to indicate the finite Trotter number $m$.
The thermal average of a physical quantity $Q$ 
is expressed as with 
the normalized eigenvector $\psi_{max}$ \cite{MS-QTMS3}
\begin{equation}\label{QTMq}
\langle Q \rangle = \lim_{m \to \infty} \, 
\langle \psi_{max} | Q | \psi_{max} \rangle_m.
\end{equation}
Suzuki et. al.\cite{MS-QTMS3} obtained a magnetization $\langle \sigma^x \rangle$ and  static real-directional correlations
 $\langle \sigma^x_i \sigma^x_j \rangle $ of the present model.

The main idea to study auto-correlation functions is follows. 
Consider the Trotter-directional correlation functions 
of the ST-transformed classical Ising spins in the extra dimension
\begin{equation}
\langle S_0S_r\rangle_m 
= \langle \psi_{max}| S_0S_r |\psi_{max} \rangle_m 
\end{equation}
where $r$ is a Trotter-directional distance of two spins.
This correlation function is also 
expressed with the original quantum spin $\sigma$ as
\begin{eqnarray}\label{mtb}
\langle S_o S_r \rangle_m &&
= \Tr e^{-\beta {\cal H}(m-r)/m} \sigma_i^z 
e^{-\beta {\cal H}r/m} \sigma_i^z /Z_m
\nonumber\\
&&{\longrightarrow} \langle \sigma_i^z(\tau)\sigma_i^z \rangle 
\qquad (\hbox{ as }  {m \to \infty})
\end{eqnarray}
with
\begin{equation}\label{tau}
 \sigma^\alpha_i(\tau)= e^{\tau {\cal H}}\sigma^\alpha_i  e^{-\tau {\cal H}}
 ,\quad  (\alpha=x, y, z)
 \quad \hbox{and} \quad \tau = \beta r/m .
\end{equation}
Thus the Trotter-directional correlation function can be interpreted as 
a Matsubara's temperature Green function \cite{Abr} after taking
the limit of the Trotter number $m \to \infty$.
The auto-correlation function is obtained by analytic continuation
of $\tau$ to imaginary time $ i t$ \cite{Abr}.

We propose the Trotter-directional correlation function 
as a quantitative and concrete measurement tool 
of quantum fluctuations.
A quantum state is usually described as a super position of classical
states such that a two spins singlet state is defined as a classical 
$|\uparrow \downarrow \rangle $ state minus $|\downarrow\uparrow \rangle$ state.
In the ST-transformed system 
we consider that the classical Ising states are stacked along 
to the extra dimension.  
The original quantum spin state is represented as a superposition of 
these stacked Ising spin states as shown in fig.\ref{fig-STtr}.
When all Ising spins have the same states, i.e.,
$\langle S_0 S_r \rangle =1$,  
the quantum fluctuation is zero, 
otherwise the quantum fluctuation exists.
If $\langle S_0 S_r \rangle =0$, the fluctuation is maximum.
Thus the correlation function can be a measure of the 
quantum fluctuation.

We study $r$ (or $\tau$) and temperature dependencies of our
correlation function in the \S 4
and we show it works well as a measure of the quantum fluctuation.
We re-derive auto-correlation functions by the QTM method in \S 5 
from the results in the \S 4 and compare with the known result
 \cite{MuSh, Perk09, Perk84, Niem, Sach, BrJa}.

%

\section{QTM and the maximum eigenvalue}

The Hamiltonian of the present transverse Ising quantum chain
\cite{XY,Pf-TI} is defined as
\begin{equation}\label{Ham}
-\beta {\cal H} = \sum_i^N \Big( K \sigma_i^z \sigma_{i+1}^z +\gamma \sigma_i^x \Big), \quad K=\beta J ,\, \gamma= \beta \Gamma.
\end{equation}

By the ST-transformation we have a two-dimensional Ising model shown in fig.\ref{fig-STtr}.
Its partition function is \cite{MS-QTM1,MI,MS-QTMS3}
\begin{equation}\label{ZN}
Z_m = A_m^{Nm} {\rm Tr} \exp \Big[ \sum_{i}^{N}\sum_{j}^{m} 
\frac{K}{m} S_{i,j} S_{i+1,j}
+\gamma'_m S_{i,j}S_{i,j+1} \Big]
\end{equation}
where 
\begin{equation}\label{param}
\exp(-2\gamma'_m) = \tanh \frac{\gamma}{m}, \qquad
 A_m=\Big( \frac{1}{2} \sinh \frac{2\gamma}{m} \Big)^{1/2} .
\end{equation}
%
The system length $N$ is assumed to be infinite. The
periodic boundary condition is required for the Trotter direction.
%
%
We can see that the vertical (horizontal) interaction
 $\gamma_m'$($\frac{K}{m}$) increases (decreases)
as the temperature increases. 

We apply the exact solution of the two-dimensional Ising model obtained by Schultz et. al.\cite{SML-TM2DI} using the 
transfer matrix method \cite{TM} to 
the present model. 
The QTM ${\cal T}$ is defined in a symmetric (and then hermitian) way 
as follows \cite{MI,MS-QTMS3}.
\begin{eqnarray}\label{TM}
&&{\cal T} = V_2^{1/2} V_1 V_2^{1/2} \quad \hbox{ with}
\nonumber\\
&&V_1 = \Big( 2\sinh \frac{2K}{m} \Big)^{m/2} \exp \big( K'_m \sum_{j}^m S_j^z \big), \quad
V_2 = \exp( \gamma'_m \sum_j^m S_j^x S_{j+1}^x ) ,
\nonumber\\
&&  \exp(-2K_m') = \tanh \frac{K}{m} .
\end{eqnarray}
Here the spin operators  $S^x,S^z$ are used. 
The original quantum spin $\sigma^z_i$ is mapped to 
the Ising spin $S_{i,j}$ in eq.(\ref{ZN}) and 
is remapped to $S^x$ in their formulation \cite{SML-TM2DI}.
Then the correlation we want to evaluate is given by 
$\langle S_0 S_r\rangle =\langle S_0^xS_r^x\rangle$ and it is
a $z$-component correlation function of the original quantum chain
as given in eq.(\ref{mtb}).

The maximum eigenvalue $\lambda_{max}$ is given by\cite{SML-TM2DI}
\begin{eqnarray}\label{eigval}
&&\lambda_{max}=\Big( 2\sinh \frac{2K}{m} \Big)^{m/2} 
    \exp\big({\frac{1}{2} \sum_q \varepsilon_{q}} \big) ,
\nonumber\\
&& \cosh \varepsilon_q = \coth\frac{2\gamma}{m}\coth \frac{2K}{m}
- \hbox{cosec}  \frac{2\gamma}{m} \hbox{cosec} \frac{2K}{m} \cos q ,
\nonumber\\
&& q= \pm \frac{1}{m}, \pm \frac{3}{m}, \cdots , \pm \frac{m-1}{m} ,
\end{eqnarray}
with the help of Jordan-Wigner, Fourier transformation and Bogoliubov diagonalization. The associated maximum eigenvector is a
vacuum of fermions.

We note that the critical line of the ST-transformed two-dimensional 
Ising model in eq.(\ref{ZN}) is given by
\begin{equation}\label{cl}
1=\sinh\frac{2K}{m} \sinh 2\gamma_m' =\sinh\frac{2K}{m}/ \sinh \frac{2\gamma}{m}
\end{equation}
as the ordinary two-dimensional Ising model.
This means that our system is critical when $\gamma=K$ 
without the dependence of the Trotter number $m$ \cite{IM}. 
When $\gamma >K$ the system is disordered and when $\gamma<K$ it is ordered \cite{Pf-TI, Sach}.

\section{Trotter-directional correlation functions}

The correlation function $\langle S_0 S_r \rangle_m$ 
 at a finite temperature 
is expressed as an $r\times r$ Toeplitz determinant \cite{SML-TM2DI,Pf-TI,XY}.
\begin{eqnarray}\label{sxx}
\langle S_0 S_r \rangle_m&&
=\langle\psi_{max} | S_0^x S_r^x | \psi_{max}\rangle_m
\nonumber\\
&&
=\left|
  \begin{array}{lllll}
    g_0   &  g_{-1}  & g_{-2}   &   &  g_{1-r}  \\
    g_1   &  g_0  &  g_{-1}  &  \cdots  &  g_{2-r}  \\
    g_2   &  g_1  &  g_0  &    &  g_{3-r}  \\
       & \vdots &     & \ddots   &    \\
   g_{r-1}    & g_{r-2}   & g_{r-3}   &  \cdots  &  g_0  \\
  \end{array}
\right|
\end{eqnarray}
It becomes zz-correlation function of the original quantum Transverse
Ising chain : $\langle \sigma^z(\tau) \sigma^z \rangle 
= \lim_{m \to \infty} \langle S_0 S_r \rangle_m$.
The matrix element $g_k$ is an expectation value 
of two fermions apart $k+1$ position and it is denoted as 
$a_{ij}$ in the reference \cite{SML-TM2DI}.
\begin{eqnarray}\label{gk}
&&
g_{\pm k} = \frac{(-1)^{k}}{m} \sum_q e^{-ikq} 
\Big[ \frac{1-e^{iq} e^{-2(\gamma+K)/m}}{1-e^{-iq} e^{-2(\gamma+K)/m}}
\frac{1-e^{-iq} e^{2(\gamma-K)/m}}{1-e^{iq} e^{2(\gamma-K)/m}} \Big] ^{1/2}
\nonumber\\
&&
=\frac{(-1)^{k}}{m} \sum_{q>0}
\frac{2\cosh(2K/m) \cos kq -e^{2\gamma/m} \cos(k+1)q -e^{-2\gamma/m} \cos(k-1)q}
{\big[ (\cosh(2\gamma/m)\cosh(2K/m) -\cos q)^2- (\sinh(2\gamma/m)\sinh(2K/m))^2 \big]^{1/2}}
\nonumber\\
\end{eqnarray}
for $K>\gamma$.
With the help of an integral formula
\begin{equation}\label{fr}
\frac{1}{\pi} \int_0^\pi \frac{1}{a+b \cos\theta} d\theta =\frac{1}{\sqrt{a^2-b^2}} \quad (a>b) ,
\end{equation}
$g_{\pm k}$ becomes
\begin{equation}\label{gk2}
g_{\pm k} = \frac{(-1)^{k}}{m \pi} \sum_{q>0}
\int_0^\pi 
\frac{2\cosh(2K/m)\cos kq -e^{2\gamma/m} \cos(k+1)q -e^{-2\gamma/m} \cos(k-1)q}
{\cosh(2\gamma/m)\cosh(2K/m)+\sinh(2\gamma/m)\sinh(2K/m)\cos\theta -\cos q} d\theta .
\end{equation}
The integrand can be simplified by using formulae of trigonometric functions 
and $\sum_{q>0} \cos nq =0 (n\neq0)$.
With the following  $s$ and $s'$ defined as
\begin{eqnarray}\label{ssd}
&&
\cosh s = \cosh\frac{2\gamma}{m} \cosh\frac{2K}{m}+\sinh\frac{2\gamma}{m} \sinh\frac{2K}{m} \cos\theta,\nonumber\\
&&s' \sinh s = \sinh\frac{2\gamma}{m} \cosh\frac{2K}{m}+\cosh\frac{2\gamma}{m} \sinh\frac{2K}{m} \cos\theta,
\end{eqnarray}
the integrand becomes
\begin{eqnarray}
&&
m \sinh\frac{2\gamma}{m} \Big( s' \sinh ks+\cosh ks \Big)
+\sum_{q>0} \frac{2\sinh(2\gamma/m)}{\cos q -\cosh s} \sinh s \Big(s'\cosh ks+\sinh ks \Big)
\nonumber\\
&&=
m \frac{\sinh(2\gamma/m)}{\cosh(m s/2)}  
\Big( -s'\sinh(\frac{m}{2}-k )s +\cosh (\frac{m}{2}-k)s \Big).
\end{eqnarray}
Here we have used the following formula
\begin{equation}\label{fr2}
\sum_{q>0}\frac{1}{\cos q -\cosh s} = -\frac{m}{2} \frac{\tanh(ms/2)}{\sinh s}.
\end{equation}
Thus we have the matrix elements
\begin{eqnarray}\label{gkf}
&&g_{\pm k}=
(-1)^{k-1} \frac{\sinh(2\gamma/m)}{\pi} \int_{0}^{\pi}
\frac{1}{\cosh (ms/2) } \Big[ s'\sinh(\frac{m}{2}-k)s  \mp \cosh( \frac{m}{2}-k)s \Big] \  d\theta
, (k\ge 1)
\nonumber\\
&&g_0=
\cosh\frac{2\gamma}{m} -\frac{\sinh(2\gamma/m)}{\pi}\int_0^{\pi}
s' \tanh \frac{ms}{2} \ \ d\theta .
\end{eqnarray}
These are valid for the both cases of $\gamma \ge K$ and $\gamma \le K$.
We note here that $g_k$ satisfies
\begin{equation}\label{pgk}
g_k(\gamma) = g_{-k}(-\gamma), \qquad 
g_{m-k}(\gamma) = (-1)^{m-1}g_{-k}(\gamma) \quad (k\ne 0).
\end{equation}
We can easily check that $\langle S_0 S_m\rangle_m=\langle S_0 S_0\rangle_m=1$ is fulfilled.

The magnetization $\langle \sigma^x\rangle$ of the original quantum spin chain is related to
the nearest neighbor correlation function $\langle S_0 S_1 \rangle_m=g_0$ 
as follows \cite{MS-QTMS3}.
\begin{equation}\label{r1cr} 
\langle \sigma^x\rangle = \lim_{m\to\infty} \coth \frac{2\gamma}{m}-g_0/\sinh \frac{2\gamma}{m}
=   \lim_{m\to\infty} \frac{1}{\pi}\int_0^{\pi} s' \tanh\frac{ms}{2} \ d\theta .
\end{equation}
For large $m$, $s$ and $s'$ in eq.(\ref{ssd}) becomes
\begin{equation}\label{sss}
s \simeq \frac{2}{m}\sqrt{\gamma^2+K^2+2\gamma K \cos\theta}, \qquad
s' \simeq \frac{\gamma+K\cos\theta}{\sqrt{\gamma^2+K^2+2\gamma K \cos\theta}},
\end{equation}
and thus we reproduce the exact solution.

The diagonal correlation function is expressed as a four-body
Ising spins correlation as follows. 
Using a Ising spin state $|S\rangle$ and its completeness $\sum_{\{S\}} |S\rangle \langle S| =1$, we obtain
\begin{eqnarray}
\langle \sigma^x(\tau)\sigma^x \rangle
&&=\lim_{m \to \infty} \sum_{\{S,S',S'',S'''\}} 
\langle S|e^{-\beta {\cal H}}e^{\tau {\cal H}}| S'\rangle
\langle S'| e^{2\gamma_m' S'S''} |S'' \rangle 
\langle S''| e^{\tau {\cal H}} |S''' \rangle 
\langle S'''| e^{2\gamma_m' S'S''} |S \rangle /Z_m 
\nonumber\\
&&
=\lim_{m \to \infty} \frac{1}{\sinh^2 (2\gamma/m)}
 \Big\langle \big(\cosh \frac{2\gamma}{m} -S_{r+1}S_r \big)
\big(\cosh \frac{2\gamma}{m} -S_{1}S_0 \big)\Big\rangle_m .
\end{eqnarray}
Here we have used an identity,
\begin{equation}
\langle S|\sigma^x e^{\gamma \sigma^x/m} |S'\rangle 
= e^{-2\gamma_m' S S'} \langle S| e^{\gamma \sigma^x/m} |S' \rangle .
\end{equation}
The 4-body correlation of Ising spins on the same Trotter axis 
is expressed by $g_k$ as
\begin{equation}\label{s4}
\langle S_0S_1S_rS_{r+1}\rangle_m = g_0^2-g_r g_{-r} .
\end{equation}
Thus we have
\begin{eqnarray}\label{xxM}
\langle \sigma^x(\tau)\sigma^x \rangle
&&= \lim_{m \to \infty}
\frac{1}{\sinh^2 (2\gamma/m)} 
\Big( \cosh^2\frac{2\gamma}{m} -2g_0^2\cosh \frac{2\gamma}{m} +
(g_0^2-g_r g_{-r}) \Big)
\nonumber\\
&&= \langle \sigma^x \rangle^2 
- \lim_{m \to \infty} \frac{1}{\sinh^2(2\gamma/m)} g_r g_{-r}.
\end{eqnarray}

Similarly, the $yy$-correlation function is given by
\begin{eqnarray}
\langle \sigma^y(\tau) \sigma^y \rangle 
&& =\lim_{m \to \infty} \frac{1}{\sinh^2 (2\gamma/m)}
 \Big\langle (- i S_r) \big( \cosh \frac{2\gamma}{m} -S_{r+1}S_r \big)
(-i S_0)\big( \cosh \frac{2\gamma}{m} -S_{1}S_0 \big)\Big\rangle_m
\nonumber\\
&&=\lim_{m \to \infty} \frac{-1}{\sinh^2 (2\gamma/m)}
 \Big[
    (\cosh^2\frac{2\gamma}{m}+1) \langle S_0 S_r \rangle_m 
     -\cosh\frac{2\gamma}{m} \big( \langle S_0 S_{r+1}  \rangle_m 
    +\langle S_1 S_r \rangle_m \big)
\Big]
\nonumber\\
&&
\end{eqnarray}
and which satisfies the following identity \cite{LaPf}
for the present model at the limit $m \to \infty$.
\begin{equation}
\frac{d^2}{d \tau^2} \langle \sigma^z(\tau) \sigma^z \rangle 
=\Gamma^2 \langle \sigma^y(\tau) \sigma^y \rangle .
\end{equation}

\section{Evaluation of the $\left\langle S_0 S_r \right\rangle_m$
correlation functions}

We study the $r$ and the temperature dependencies of the correlation 
functions $\langle S_0 S_r \rangle_m$ in this section.
We cannot apply Szeg\"o's theorem \cite{McWu} to our Toeplitz determinant in 
eq.(\ref{sxx}). If we 
take $m\to\infty$ for a fixed $r$, all $g_k$ becomes $0$ except
 $g_0 \to 1$ and thus
we always have a wrong result $\langle S_0S_r\rangle_m=1$. 
We need to evaluate the determinant with a finite $m$
and then take $m \to \infty$ limit at final to get a
correct result of the original quantum chain \cite{MI}.

Define $\delta=r/m$ to indicate the normalized distance of two spins. 
We assume $m$ and $r$ are very large but finite. 
We take a perturbative approach
for the small $\delta$ in the first subsection.
We do numerical computation
for the general $\delta$ in the successive subsections.

\subsection{Analytic approach for the small $\delta$}

We calculate the determinant for the small $\delta$ up to 
the second order analytically. 
Expand the matrix elements $g_k$ up to $k^2$ order, 
\begin{eqnarray}\label{gks}
&&g_{\pm k} \simeq
(-1)^k \sinh\frac{2\gamma}{m} \Big[
\pm 1- \frac{1}{\pi}\int_{0}^{\pi} s' \tanh\frac{ms}{2} \ d\theta
\nonumber\\
&& \qquad
+ k \frac{1}{\pi}\int_{0}^{\pi}s(s'\mp \tanh\frac{ms}{2} ) \ d\theta
+\frac{k^2}{2} \frac{1}{\pi}\int_{0}^{\pi} s^2 (\pm 1 - s' \tanh \frac{ms}{2} \ d\theta
\Big].
\end{eqnarray}
In general the determinant of a Toeplitz matrix 
whose elements has up to $k^2$ dependence can be
obtained with the help of the standard matrix procedure and 
is given in Appendix A.

After tedious but straightforward calculation we have
\begin{eqnarray}\label{sxx-small}
\langle  S_0 S_r \rangle_m&&=1+ c_1(m) \delta+ c_2(m) \delta^2 +O(\delta^3) \nonumber\\
\hbox{with}&&
c_1(m)=-2\gamma I_0-\frac{4\gamma}{3 m^2} \big( 2\gamma^2 I_0 -2\gamma I_1 +I_3 \big)+O(m^{-4}),
\nonumber\\
&&
c_2(m)=2\gamma^2-\frac{4\gamma^2}{3m^2} \big(I_1^2-I_0 I_2 +K^2 \big)+O(m^{-4}).
\end{eqnarray}
Where integrals are defined as
\begin{eqnarray}\label{intp}
&&I_0=\frac{1}{\pi}\int_{0}^{\pi}
\frac{\gamma+K \cos\theta}{\sqrt{\gamma^2+K^2+2\gamma K \cos\theta}} \tanh\sqrt{\gamma^2+K^2+2\gamma K \cos\theta} \ d\theta
= \langle \sigma^x\rangle ,
\nonumber\\
&&I_1=\frac{1}{\pi}\int_{0}^{\pi}
\sqrt{\gamma^2+K^2+2\gamma K \cos\theta} \tanh\sqrt{\gamma^2+K^2+2\gamma K \cos\theta} \ d\theta ,
\nonumber\\
&&I_2=\frac{1}{\pi}\int_{0}^{\pi}
(\gamma+K \cos\theta)\sqrt{\gamma^2+K^2+2\gamma K \cos\theta} \tanh\sqrt{\gamma^2+K^2+2\gamma K \cos\theta} \ d\theta ,
\nonumber\\
&&I_3=\frac{1}{\pi}\int_{0}^{\pi}  \Big(
\frac{\gamma K^2(\gamma^2+2K^2+3\gamma K \cos\theta)\sin^2\theta}{
\sqrt{\gamma^2+K^2+2\gamma K \cos\theta}^3} \tanh\sqrt{\gamma^2+K^2+2\gamma K \cos\theta} 
\nonumber\\
&&\qquad \qquad \qquad
+\frac{\gamma^2K^2(\gamma+K \cos\theta)\sin^2\theta}{(\gamma^2+K^2+2\gamma K \cos\theta)\cosh^2\sqrt{\gamma^2+K^2+2\gamma K \cos\theta}} \Big) \ d\theta .
\end{eqnarray}
The first integral $I_0$ is nothing but the magnetization 
$\langle \sigma^x\rangle=-\frac{\partial}{\partial \gamma} ( \beta f) $ of the original quantum chain.
It takes a finite value between 0 and 1 depending on the ratio $K/\gamma$ 
in the low temperature region and 
it becomes $\gamma$ at the high temperature limit.

We can check easily that the result eq.(\ref{sxx-small}) agrees with a perturbative calculation of
the temperature Green function for a small $\tau=\beta \delta $ ,
\begin{equation}
\langle \sigma^z(\tau)\sigma^z \rangle
=
\Big\langle \big(\sigma^z+ [{\cal H},\sigma^z ] \tau 
+\frac{1}{2} [{\cal H},[{\cal H},  \sigma^z]] \tau^2
+\cdots \big) \sigma^z \Big\rangle 
= 1- 2\Gamma  \langle \sigma^x \rangle \tau + 2\Gamma^2 \tau^2 +\cdots .
\end{equation}

\subsection{Numeric evaluation for $K/\gamma =0.7$ in the disordered region.}

To study the behavior of our correlation functions for the whole 
range of the parameters $\delta$ and the temperature we perform
 a direct numerical computation
of the integrals of eq.(\ref{gkf}) and the determinant of eq.(\ref{sxx}).
The data are taken for $K/\gamma=0.7,1.0,1.3, 2.0$, 
$m=8 \sim 1024$
and $\gamma= 0.1 \sim 64$ while $\gamma/m<1$. The distance 
$r$ varies $r=1,2,3,\cdots, m/2,m/2+1$.
Due to the periodicity for the Trotter direction, 
our correlation function
has the same value at $r$ and at $m-r$.
It is a good test for our numeric computation whether the
function has the same value at $r=m/2-1$ and $r=m/2+1$
or not.
Our data has 9 digits accuracy even in the worst case.

\begin{figure}[tbp]
  \begin{center}
    \includegraphics[keepaspectratio=true,height=60mm]{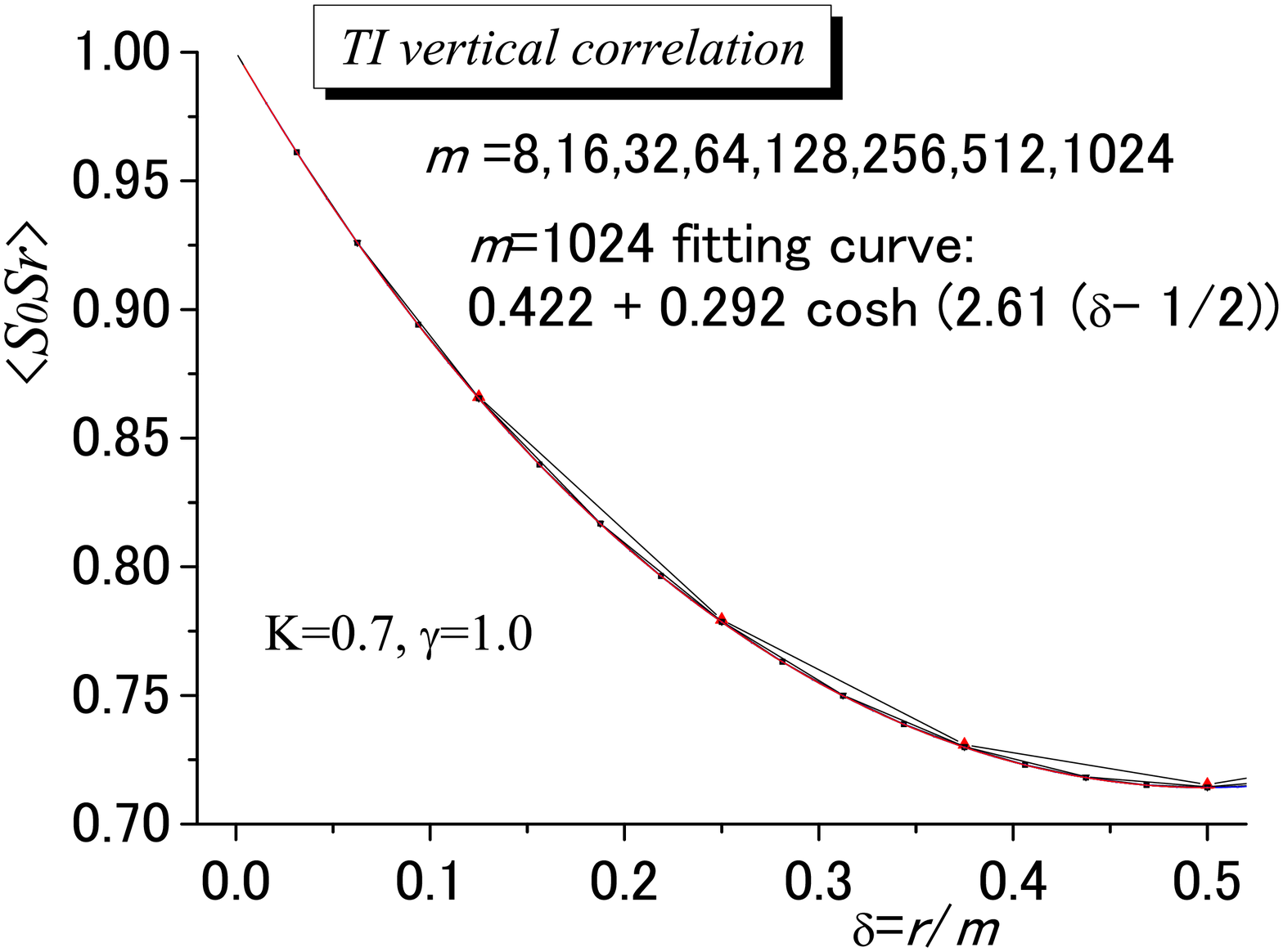}
  \end{center}
  \caption{$\langle S_0 S_r \rangle_m$ versus $\delta=r/m$ at $K=0.7$
  and $\gamma=1.0$ for various $m=8 \sim 1024$.
  The fitting curve for $m=1024$ is also drawn.}
  \label{fig-07KGHighT}
\end{figure}

Figure \ref{fig-07KGHighT}  is the $\delta$ dependence of the 
correlation function
for the high temperature $K=0.7$ and $\gamma=1$. 
We omit the data for $\delta>1/2$.
It takes the lowest value
at $\delta=1/2$.
Its $m$-dependence is very week as was shown in the analytic evaluation 
in eq.(\ref{sxx-small}) ($m^{-2}$ dependence) so that 
we can not distinguish each curves of the different $m$ in this figure.

We adopt the following two parameters ($b$ and $c$) fitting curve
to analyze the $\delta$ dependence,
\begin{equation}\label{regr}
f_{H}(\delta)=1-\Big( \cosh(c/2)-\cosh (c(1/2-\delta))\Big)/\cosh b
\end{equation}
which satisfies $f_H(0)=1$.
This curve is exact for $K=0$ as shown in Appendix B. 
\begin{equation}\label{regK0}
\langle S_0 S_r \rangle_m=\cosh(\gamma(1-2\delta))/\cosh \gamma.
\qquad (c=2\gamma,\,\, b=\gamma)
\end{equation}
The fitting gives excellently good agreement to the numerical data
as shown in fig.\ref{fig-07KGHighT}.
%
%
%
%
%

%
\begin{figure}[tbp]
  \begin{center}
    \includegraphics[keepaspectratio=true,height=60mm]{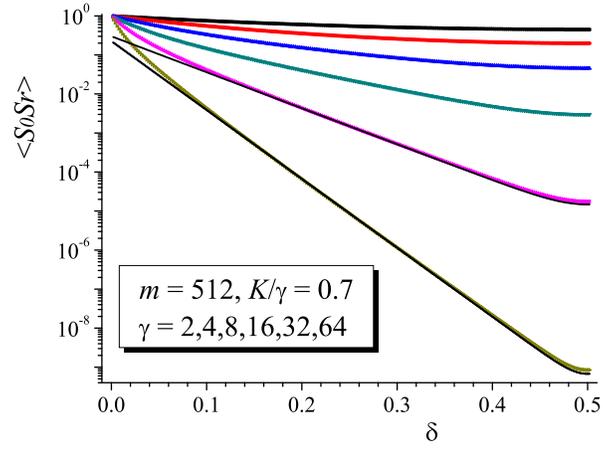}
  \end{center}
  \caption{Semi-log plot of correlation functions versus $\delta$ for
   $\gamma=2, 4, 8, 16, 32, 64$ from top to bottom. 
   $K/\gamma=0.7$ and $m=512$. 
   Two thin lines are fitting curves of eq.(\ref{reg3}).}
  \label{fig-07KGLowT-r}
\end{figure}

At a low temperature the correlation function shows exponential decay
as shown in fig.\ref{fig-07KGLowT-r}. 
We use a new fitting curve  
which does not satisfy $f_L(0)=1$
any more to estimate the main term.
\begin{equation}\label{reg2}
f_L(\delta)= \cosh(c(1/2-\delta)) / \cosh b .
\end{equation}
The results are
\begin{eqnarray}\label{reg3}
&&
\cosh(21.2(1/2-\delta))/\cosh 11.8 \quad \hbox{ for } \quad \gamma=32,
\nonumber\\
&&
\cosh(40.6(1/2-\delta))/\cosh 21.8 \quad \hbox{ for } \quad \gamma=64,
\end{eqnarray}
and are drawn in fig. \ref{fig-07KGLowT-r}. The values of $b$ and $c$ for
$\gamma=64$ are about twice larger than those for $\gamma=32$, respectively.

\begin{figure}[tbp]
  \begin{center}
    \includegraphics[keepaspectratio=true,height=60mm]{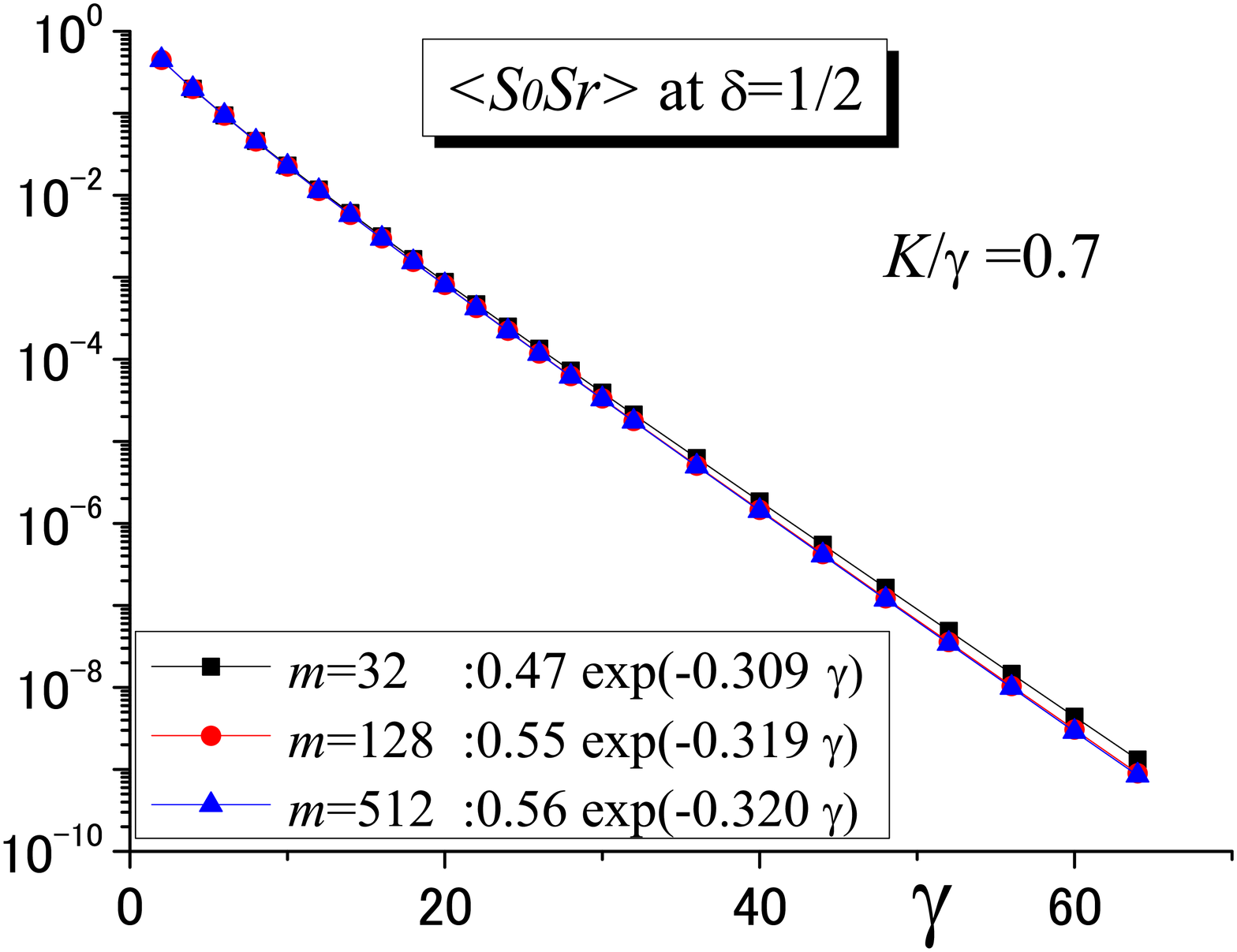}
  \end{center}
  \caption{Semi-log plot of the correlation function at $\delta=1/2$ versus 
  $\gamma$. $K/\gamma=0.7$. $m=32, 128,512$. Lines are fitting curves shown in the legend. }
  \label{fig-07KGLowT-beta}
\end{figure}

To study the temperature dependence we plot the data for $\delta=1/2$ in
 fig.\ref{fig-07KGLowT-beta}. The exponential decay for the $\gamma$ is
 clearly shown. By fitting the exponential function 
to the data of $m=32,128,512$,
 we estimate the gradient approximately 
\begin{equation}
\exp(-0.32 \gamma).
\end{equation}

From these results we conclude that the correlation function decays 
exponentially to zero 
for both the $\delta$ and the temperature ($\gamma$) 
in the disordered region.

\subsection{Critical region: $K=\gamma$}

\begin{figure}[tbp]
  \begin{center}
    \includegraphics[keepaspectratio=true,height=60mm]{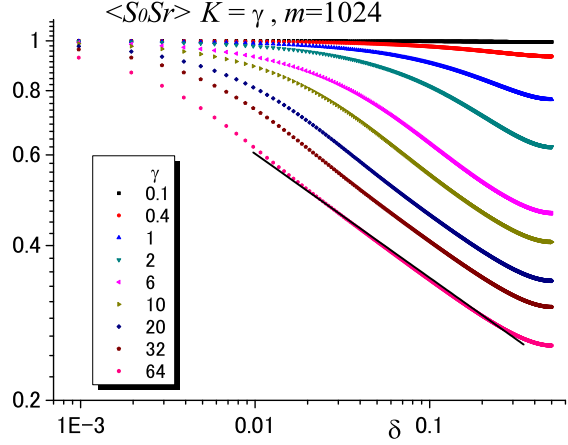}
  \end{center}
  \caption{Log-log plot versus $\delta$ 
  for various $\gamma$. $K=\gamma$ and $m=1024$.
  The line for $\gamma=64$ is a fitting curve given in eq.(\ref{eta}).}
  \label{fig-KeqG-r}
\end{figure}

\begin{figure}[tbp]
  \begin{center}
    \includegraphics[keepaspectratio=true,height=60mm]{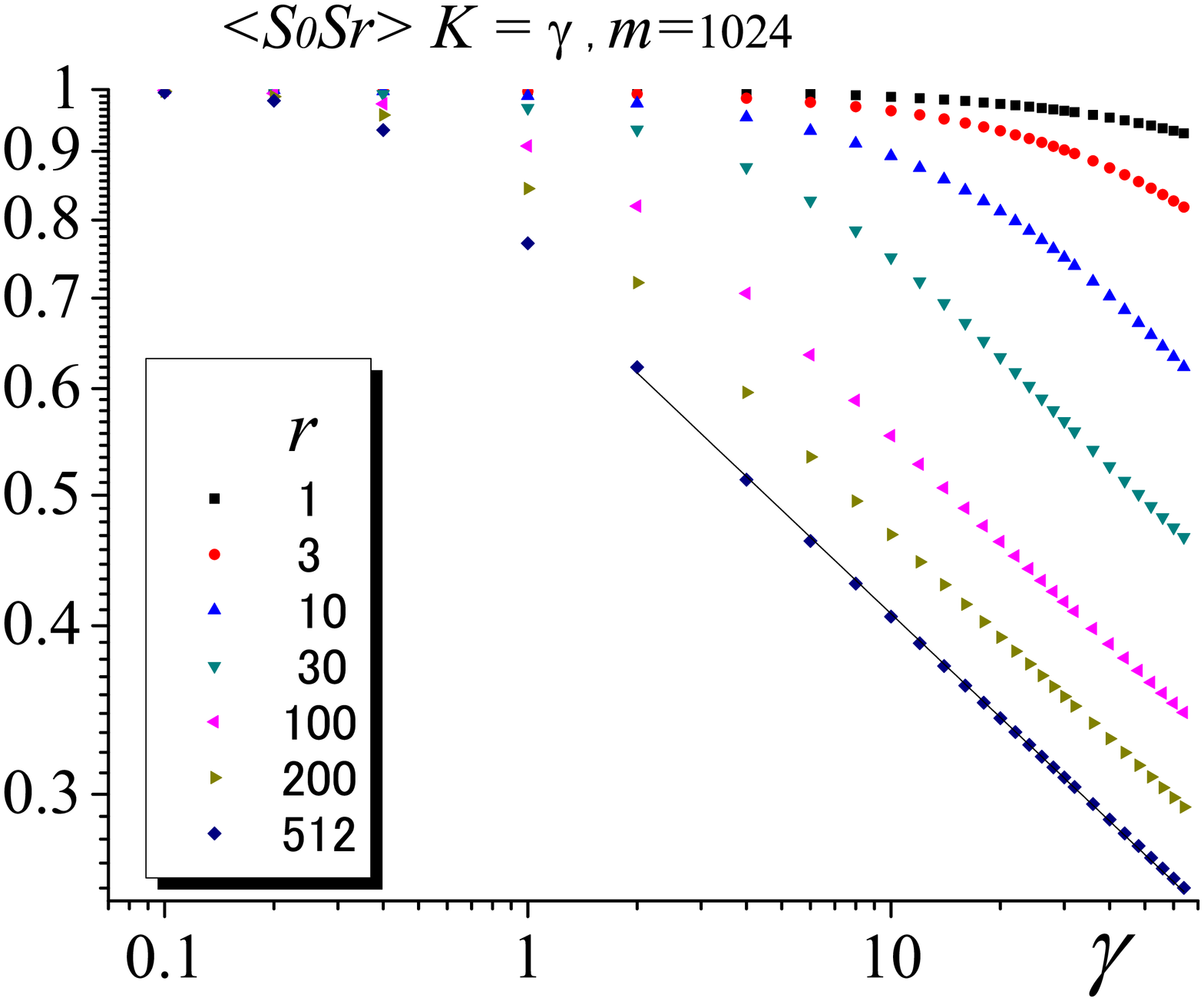}
  \end{center}
  \caption{Log-log plot versus $\gamma$ for various position $r$. 
  The line is a fitting curve for $r=512$ given in eq.(\ref{reg-KeqG-r}).}
  \label{fig-KeqG-beta}
\end{figure}

We show two graphs figs. \ref{fig-KeqG-r} and  \ref{fig-KeqG-beta}
for $m=1024$. The former shows the $\delta$-dependence and the
latter shows the $\gamma$-dependence. 
The both graphs clearly show a power law behavior of the correlation functions
in the low temperature.
The fitting line for $\gamma=64$ in fig.\ref{fig-KeqG-r} 
is
\begin{equation}\label{eta}
1.01 \times r ^{-0.24}.
\end{equation}
The fitting line shown in fig.\ref{fig-KeqG-beta} for 
$r=512$ ($\delta=1/2$) is
\begin{equation}\label{reg-KeqG-r}
0.72 \times \gamma^{-0.25}.
\end{equation}
These exponents remind us the two-dimensional Ising model which
has $\eta=1/4$ and $\beta=1/8$.




\subsection{Ordered region: $K/\gamma=1.3$}

\begin{figure}[btp]
  \begin{center}
    \includegraphics[keepaspectratio=true,height=60mm]{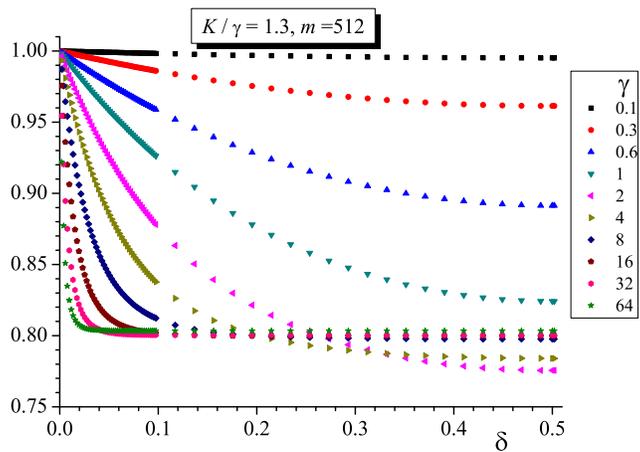}
  \end{center}
  \caption{Normal plot of $m=512$ for the several temperatures.}
  \label{fig-13KG-r}
\end{figure}

\begin{figure}[tbp]
  \begin{center}
    \includegraphics[keepaspectratio=true,height=60mm]{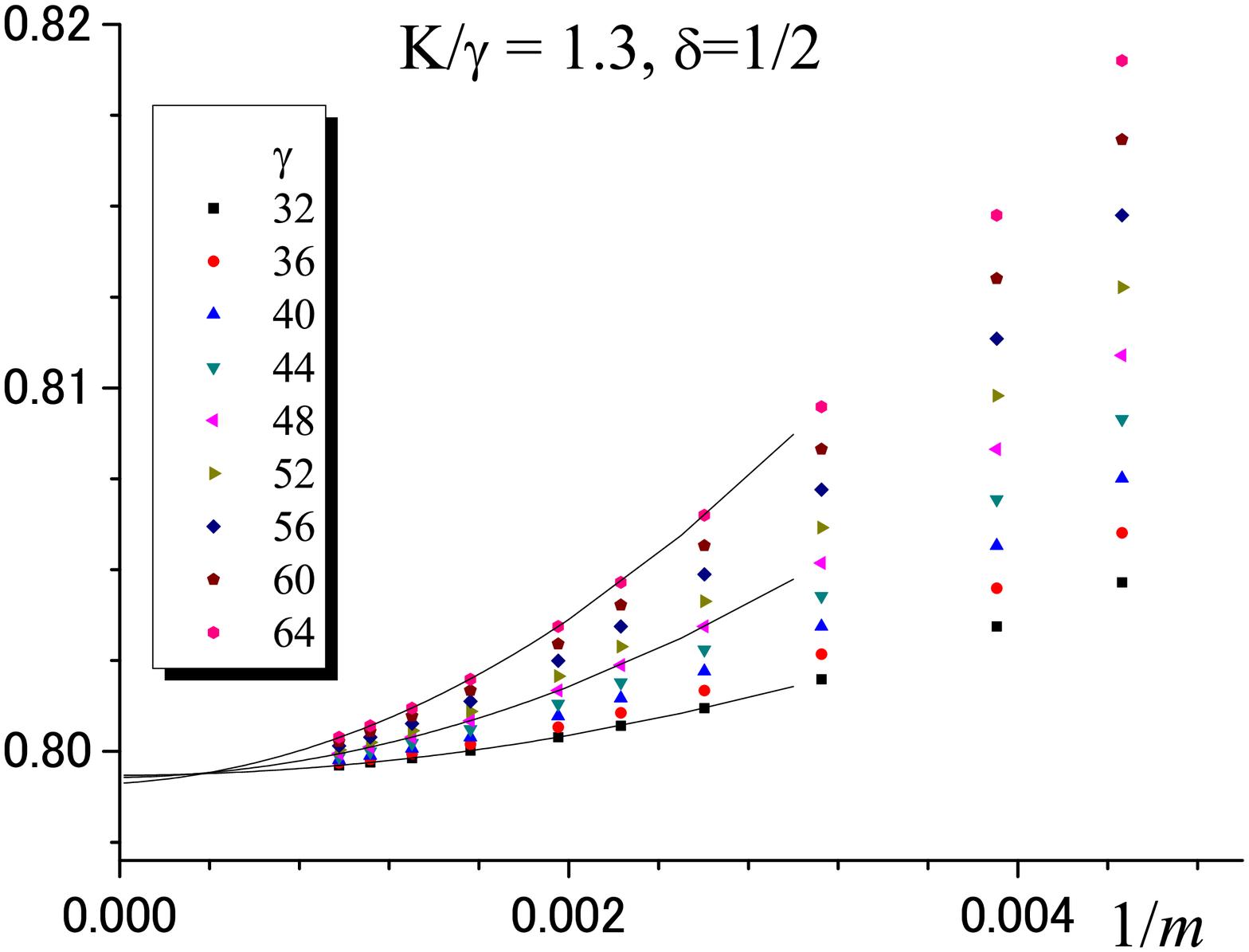}
  \end{center}
  \caption{Saturation values versus $m^{-1}$ for $K/\gamma=1.3$. 
$\gamma=64 \sim 32$ from top to bottom.
$m=224, 256, 320, 384, 448, 512, 640, 768, 896, 1024$. Fitting curves are quadratic equations of $1/m$.}
  \label{fig-13KGsatur-m}
\end{figure}

The big difference from the other regions is that 
the correlation function decreases and saturates to some
value as the temperature decreases in the ordered region.
In fig.\ref{fig-13KG-r} the saturated
value is about 0.8 for $K/\gamma=1.3$.
This means the quantum fluctuation is small in this region.
The increase  for the large $\delta$ and the low temperature
($\gamma>2$) is due to the finite $m$.
As is seen in fig.\ref{fig-13KGsatur-m}, this increase
disappears as $m \to \infty$.

We estimate the saturation values as follows.
Plot the data of $\delta=1/2$ versus $1/m$
in fig.\ref{fig-13KGsatur-m}. Fit them with a quadratic equation
of $1/m$ and find the limit values of $m \to \infty$  for each temperature. 
 Three these fitting curves are also drawn in fig.\ref{fig-13KGsatur-m}.
The three limit values almost coincide to each others such that
$0.79934$ for  $\gamma=32$, $0.79928$ for $\gamma=48$ and 
$0.79912$ for $\gamma=64$. Thus we conclude the saturation value
for $K/\gamma=1.3$ at the zero temperature is about 0.799.
Additionally, when $K/\gamma=2.0$ the saturation value is estimated 
about 0.930. 

These values are equal to the square of the order parameter $\langle \sigma^z \rangle ^2 = (1-\gamma/K)^{1/4}$ at $T=0$ \cite{Pf-TI,Perk09, Sach}. 

When $\gamma=0$ ($K/\gamma=\infty$), the value is 
$\langle S_0 S_r\rangle_m=1$ for any $r$ 
since $g_{\pm k} =0$ and $g_0=1$.

\section{Auto correlation functions}
We assume that the analytic continuation from the temperature
Green function to the time-dependent auto-correlation function
is achieved by just replacing $\tau$ with the imaginary time $it$
in the present case. 
This assumption is confirmed by deriving the same result as
the known results \cite{MuSh,Perk09,Perk84,Niem,BrJa,Sach}.

\subsection{$\langle \sigma^z(t) \sigma^z \rangle$}
When $t \simeq 0$, replacing $\delta \to \tau/\beta \to it/\beta$ in
eq.(\ref{sxx-small}) for $m=\infty$,
\begin{equation}
\langle \sigma^z(t) \sigma^z \rangle 
=\lim_{m \to \infty} \langle S_0 S_r \rangle_m 
=1- 2i \Gamma \langle \sigma^x \rangle t - 2\Gamma^2 t^2 +O(t^3)
\end{equation}
This result agrees to  
Brandt and Jacoby\cite{BrJa} and Perk et.al.\cite{Perk84} with 
rescaling $t$ by $t/2$ due to the definition of the Hamiltonian.

\begin{figure}[tbp]
  \begin{center}
    \includegraphics[keepaspectratio=true,height=80mm]{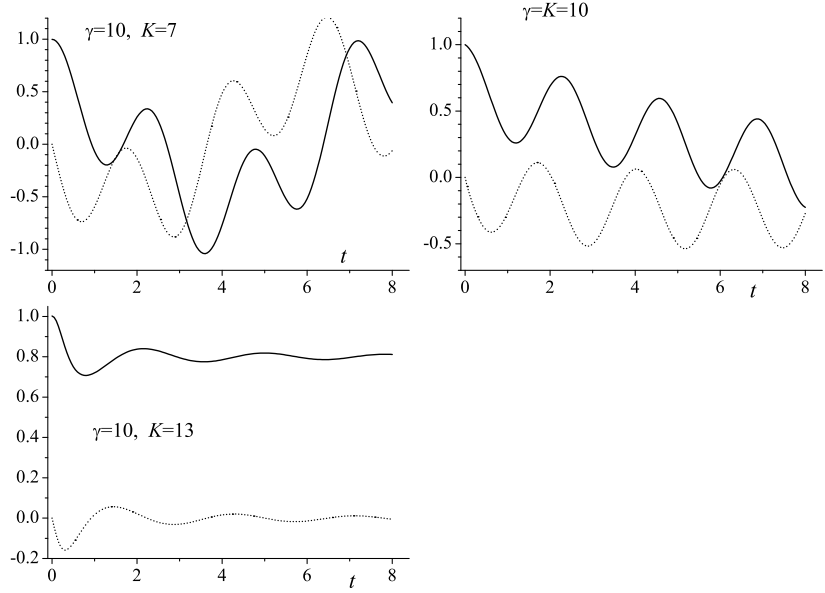}
  \end{center}
  \caption{Time dependence of the auto-correlation functions
  $\langle \sigma^z(t)\sigma^z \rangle$.
  Solid lines are real part and dotted lines are imaginary part
  of the functions.
  The graph of $\gamma=10,K=7$ should be compared to figs. 3 and 4
  of Perk and Au-Yang \cite{Perk09}, and the graph of $\gamma=10,K=13$ 
  to figs. 6 and 7 of it.}
  \label{fig-ac}
\end{figure}

We adopt the following functions for the general $t$
with many fitting parameters to
fit whole the range $0 \le \delta \le 1/2$ of 
the numerical data shown in 
figs.\ref{fig-07KGLowT-r},\ref{fig-KeqG-r} and \ref{fig-13KG-r}.
Each fitting function is chosen to have the main exponential or power decay terms found in \S4.
$\delta$ is replaced  to $\tau=\beta \delta$ ($\beta=\gamma$).
The used data are for $m=512$ and
$\gamma=10$.
The results are 
\begin{equation}
0.028+0.48e^{-2.7 \tau }+0.64 e^{-0.74\tau}
+\frac{3.1 e^{-0.44\tau}}{\tau-20.} \qquad \hbox{for } \gamma=10, K=7,
\end{equation}
\begin{equation}
-2.9 +8.3(1-1.0e^{0.041\tau}) \tau^{-1/4}+0.30e^{-2.7\tau}
+3.7e^{0.048\tau}  \qquad \hbox{for } \gamma=K=10,
\end{equation}
\begin{equation}
0.85+\frac{0.088e^{-2.2\tau}}{0.44 +\tau}
-\frac{270e^{0.032\tau}}{(70 +\tau)^2}
+\frac{4700e^{-2.5\tau}}{(400 +\tau)^3}  \qquad 
\hbox{for } \gamma=10, K=13
\end{equation}
and are shown in figs.\ref{fig-ac} after replacing $\tau$ to $it$.
These shows almost the same characteristic behavior obtained by 
many authors \cite{Perk09,BrJa,MuSh,Sach}
such that a fast+slow oscillating decay around zero for $\gamma>K$, 
a simple oscillating decay for $\gamma=K$ and
a saturation to a non-zero value with a oscillation for $\gamma<K$.
It is, as is well known, not easy to get accurate results
 by an analytic continuation from numerical data. 
%
%
We need to consider $m$ dependence to get more accurate result as
has done in \S 4 or need a more sophisticated method. 
It is instructive that the Trotter number $m=512$ is
large enough to estimate the leading singularity as shown in \S 4, however,
it is small to estimate the whole range of the time dependence 
of the auto-correlations quantitatively. 
The present result is an example obtained by a simple extrapolation.

\subsection{$\langle \sigma^x(t) \sigma^x \rangle$}

Substitute $g_r$ and $\tau=it = \beta r/m$ into eq.(\ref{xxM}) 
and take the limit $m \to \infty$
we have
\begin{eqnarray}\label{acxx}
&&
\langle \sigma^x(t) \sigma^x \rangle
=
\langle \sigma^x \rangle^2 
- \bigg[ \frac{1}{\pi} \int_{0}^{\pi}
 \frac{\Gamma+J\cos\theta}{w}
    \Big( \tanh\beta w
        \cosh(2iwt)
        -\sinh(2iwt) \Big)d \theta \bigg]^2
\nonumber \\
&&
\qquad \qquad\qquad
+\bigg[\frac{1}{\pi} \int_{0}^{\pi}
    \Big(\cosh(2iwt)-\tanh\beta w \sinh(2iwt) \Big)
    d\theta
\bigg]^2
\end{eqnarray}
for a finite temperature
where we have abbreviated
$ w = \sqrt{\Gamma^2+J^2+2\Gamma J \cos \theta}$.
Thus we have obtained the Niemeijer's exact result \cite{Niem,MaSi,MuSh} by
extending the QTM method.

\section{Summary and Discussions}
We have studied the Trotter-directional correlation function
which is interpreted as a Matsubara's temperature Green function
by the Quantum Transfer Matrix method.

By applying analytic continuation to the correlation function
we examined the auto-correlation
functions $\langle \sigma^z(t) \sigma^z\rangle$ 
and $\langle \sigma^x(t) \sigma^x\rangle$. 
The numerical result for
the $zz$-correlation caught the characteristic behavior of the
time dependence of them.
We have re-derived the $xx$-correlation exactly.

Our formulation are based on the ST-transformation and thus $\beta/m$ should be small enough to get a meaningful result of the relevant
quantum systems. 
In practice we use a discrete $\tau= \beta \delta= \beta r/m$ 
to estimate the continuous $\tau$ or $t$  dependence of
auto-correlations. Thus we need a large $m$ to get more
accurate results in numerical calculation.

We have demonstrated that the Trotter-directional correlation function
is a useful tool to measure the quantum fluctuations.
It has the same value of the square of the order parameter 
$\langle \sigma^z \rangle ^2$ at $T=0$. 
The shrinkage of a spin length which is given as 
the order parameter is often used as a measure of the 
quantum fluctuation.
While the order parameter $\langle \sigma^z \rangle$ is zero for $T>0$ for the 
present one-dimensional 
system and then it can not be used as a measure,
our correlation has a finite value which
varies from $\langle \sigma^z \rangle^2 (T=0)$ to 1 ($T= \infty$). 
When $\gamma=0$ which means no quantumness (1 dimensional classical Ising model)
, our correlation is equal to one 
and thus indicates no quantum fluctuation correctly.
The square of the magnetization $\langle \sigma^z \rangle ^2 \simeq 
\langle \sigma_0\sigma_R\rangle$ ($R \to \infty$) 
is considered as a projection to an another spin in distance $R$.
Our correlation
$\langle \sigma_i^z(\tau)\sigma_i^z\rangle$ is a projection of
itself and also it shows how the 
fluctuation increases as a function of "time"($\tau$). 
Thus our correlation function and the order parameter 
are complemental measures for the quantum fluctuation.

We emphasize that the Trotter-directional correlation function
can be defined for any ST-transformed
classical systems mapped from quantum systems. 
What we should observe is the classical Ising spins $\{S\}$
and thus we can easily done by the Quantum Monte Carlo simulation or others.
By using many Ising spins which are on different stacked layers
or different real positions, we can 
study general many temperature/time Green functions.

Finally we mention Bethe-Ansatz systems. The
maximum eigenvalue and the associated eigenvector of the Heisenberg chain
are well known \cite{Koma,AK-ST,Corr-XXZ} so that the application our
method to them is a future problem.

\subsection*{Appendix A: Determinant of a Toeplitz Matrix}

The determinant of the $r\times r$ Toeplitz matrix with the
elements
\begin{equation}
M_{i,j}=\left\{
  \begin{array}{cc}
    a+(i-j)b+(i-j)^2 c   &  : 1\le j<i \le r  \\
    X_0   &  :1 \le i=j \le r \\
    d+(j-i)e+(j-i)^2f   & :1\le i<j \le r   \\
  \end{array}
\right.
\end{equation}
is expressed as follows.
\begin{eqnarray}
\hbox{det}M=&&\Big(\frac{c}{c-f}\Big)^3 (A_0)^r-\Big(\frac{f}{c-f}\Big)^3 (\overline{A_0})^r
\nonumber\\
&& -\frac{(A_0)^{r-1}}{(c-f)^3} \bigg[
\Big\{ -6c^2f A_0 p(r+1) -\overline{z_3}p(r) +\overline{z_2} p(r-1) \Big\} /2\nonumber\\
&&\qquad\qquad   -r c \Big\{ (\overline{z_1} +4cf(c-f))p(r) - \overline{z_1} p(r-1) \Big\}\nonumber\\
&&\qquad\qquad   +r^2 c^2 f(c-f) \Big\{ p(r) -  p(r-1) \Big\} \bigg]\nonumber\\
&& - \frac{(A_0)^{r-2}}{(c-f)^3} \bigg[ \Big\{
6cf^2 A_0^2 q(r+1)+A_0 z_3 q(r)-\overline{A_0} z_2 q(r-1) \Big\}/2 \nonumber\\
&&\qquad\qquad
+r f \Big\{ A_0( z_1-4cf(c-f))q(r)-\overline{A_0} z_1 q(r-1)\Big\}\nonumber\\
&&\qquad\qquad  +r^2 cf^2(c-f) \Big\{A_0 q(r)-\overline{A_0} q(r-1) \Big\}
\bigg] .
\end{eqnarray}

Here
\begin{eqnarray}
&&z_1=c^2e-f^2b+2bcf+2cf(c-f), \nonumber\\
&&z_2=c^2e^2+2c(-cd+(b+c)e)f+(b^2+4bc+2c(2a-3X_0+c+3d))f^2-2(a+b+c)f^3,
\nonumber\\
&&z_3 = z_2-4f z_1+2cf^2(3a-3X_0+3b+3e+c-f) ,
\nonumber\\
&&A_0=X_0-d, \qquad A_1=-3X_0+a+b+c+2d+e-f.
\end{eqnarray}
The symbol overline $\overline{z}$ means an exchange
 $(a,b,c) \leftrightarrow (d,e,f)$
of $z$. (transpose of the matrix)
And
\begin{eqnarray}
&p(k)&=\frac{1}{(\alpha-\beta)(\beta-\gamma)(\gamma-\alpha)}
\Big[ \alpha^{k+1}(\beta-\gamma)+\beta^{k+1}(\gamma-\alpha)
+\gamma^{k+1}(\alpha-\beta) \Big], \nonumber \\
&q(k)&=\frac{1}{(\alpha-\beta)(\beta-\gamma)(\gamma-\alpha)}
\Big[ (\alpha\beta)^{k}(\alpha-\beta)+(\beta\gamma)^{k}(\beta-\gamma)
+(\alpha\gamma)^{k}(\gamma-\alpha) \Big] ,
\nonumber\\
&&
\overline{p(k)} = \Big( A_0/ \overline{A_0} \Big)^{k-1} q(k) .
\end{eqnarray}
$\alpha,\beta,\gamma$ are roots of $A_0 x^3+A_1 x^2-\overline{A_1}x-\overline{A_0}=0$ .

Multiply $(-1)^i \times (-1)^{j-1}= (-1)^{i-j-1}$ to $M_{i,j}$
to suite our matrix $M'$ with elements:
\begin{equation}
M'_{i,j}=\left\{
  \begin{array}{cc}
   (-1)^{i-j-1} \big\{ a+(i-j)b+(i-j)^2 c \big\}   &  : 1\le j<i \le r  \\
    -X_0   &  :1 \le i=j \le r \\
   (-1)^{j-i-1} \big\{ d+(j-i)e+(j-i)^2f \big\}   & :1\le i<j \le r   \\
  \end{array}
\right. .
\end{equation}
The relation $M$ and $M'$ is
\begin{equation}
\hbox{det} M' = (-1)^r \hbox{det} M .
\end{equation}

\subsection*{Appendix B: $\langle S_0S_r \rangle_m $ for $K=0$ 
(derivation of Eq.(\ref{regK0}))}

When $K=0$, the system reduces to a one-dimensional Ising chain
for the Trotter direction with a spin interaction $\gamma'_m$.
The correlation function is simply given by
\begin{equation}\label{Ap-cK0}
\langle S_0S_r \rangle_m = \frac{\tanh^r \gamma'_m +\tanh^{m-r} \gamma'_m }
{1+\tanh^m \gamma'_m}
= \frac{\cosh(\gamma(1-2\delta))}{\cosh\gamma} 
\end{equation}
under the periodic boundary condition with using eq.(\ref{TM}). 

Alternatively, we can derive this result from eq.(\ref{sxx}).
The $s$ and $s'$ of eq.(\ref{ssd}) are now
\begin{equation}\label{ssdK0}
s=\frac{2\gamma}{m}, \quad s'=1 .
\end{equation}
The elements $g_k$ are simplified as
\begin{eqnarray}\label{Ap-gk}
&&g_{\pm k} = (-1)^{k-1} \frac{\sinh(2\gamma/m)}{\cosh \gamma}
\Big[ \mp \exp \big( \mp (1-\frac{2 k}{m}) \big) \gamma \Big] ,
\nonumber\\
&&
g_0= \cosh \frac{2\gamma}{m} -\sinh \frac{2\gamma}{m} \tanh \gamma .
\end{eqnarray}
Thus the matrix of eq.(\ref{sxx}) can be diagonalized and
the determinant is obtained directly.

As the third method, we can derive the result eq.(\ref{Ap-cK0}) 
without the ST-transformation by using a Heisenberq equation:
\begin{equation}
\frac{d}{d\tau} \sigma^z(\tau)  = [ {\cal H}, \sigma^z(\tau)],
\qquad \sigma^z(\tau)=e^{{\cal H}\tau} \sigma^z e^{-{\cal H}\tau} 
\end{equation}
where $\tau$ is imaginary time and ${\cal H} = -\sum_j \Gamma \sigma_j^x$.
Using twice this equation, $\sigma^z(\tau)$ follows a
differential equation and it becomes as
\begin{equation}
\frac{d^2}{d \tau^2} \sigma^z(\tau)= \Gamma^2 \sigma^z(\tau) , \qquad
\sigma^z(\tau) = c_1 e^{2\Gamma \tau} + c_2 e^{-2\Gamma \tau}.
\end{equation}
$c_1$ and $c_2$ are constant $2\times 2$ matrices and are defined 
by an initial $\langle \sigma^z(0)\sigma^z\rangle =1$ and
a periodicity $\langle \sigma^z(\tau)\sigma^z\rangle
= \langle \sigma^z(\beta-\tau)\sigma^z\rangle$
conditions\cite{Abr}. We have  again 
\begin{equation}
 \langle \sigma^z(\tau)\sigma^z\rangle 
 = \lim_{m \to \infty} \langle S_0S_r\rangle_m 
 = \frac{\cosh(\gamma-2\Gamma\tau))}{\cosh\gamma}. \quad(\Gamma \tau= \gamma \delta) .
 \end{equation}

\end{document}